\documentclass[11pt]{article}
\begin{document}
\title
{Quantum mechanics of a charged particle in a background magnetic field interacting with linearized gravitational waves}

\author{
{\bf {\normalsize Sunandan Gangopadhyay}$^{a,b,c}
$\thanks{sunandan.gangopadhyay@gmail.com, sunandan@bose.res.in}},
{\bf {\normalsize Anirban Saha}
$^{a,c}$\thanks{anirban@iucaa.ernet.in}}\\
$^{a}$ {\normalsize Department of Physics, West Bengal State University, Barasat, India}\\
$^{b}${\normalsize Visiting Associate in S.N. Bose National Centre for Basic Sciences,}\\
{\normalsize JD Block, Sector III, Salt Lake, Kolkata 700098, India}\\
$^{c}${\normalsize Visiting Associate in Inter University Centre for Astronomy $\&$ Astrophysics,}\\
{\normalsize Pune, India}\\[0.3cm]
}
\date{}
\maketitle
\begin{abstract}
{\noindent We consider the dynamics of a charged particle interacting with background electromagnetic field under the influence of linearized gravitational waves in the long wave-length and low-velocity limit. 
Following the prescription in \cite{speli}, the system is quantized and the Hamiltonian is then solved by using standard algebraic iterative methods. The solution is in conformity with the classical analysis and 
shows the possibility of tuning the frequency by changing the magnetic field to set up resonance.} 

\end{abstract}
\maketitle
\noindent With the development of various ground based gravitational wave (GW) detectors like LIGO \cite{abramovici}, the possibility of direct detection of GW(s) with a strain sensitivity of the order of $h \sim 10^{-21}/\sqrt{Hz} $ or better in the frequency range between 100-1000 Hz is expected in near future \cite{thorn}. With such a small response of matter to a passing GW, more often than not, a quantum mechanical treatment of this interaction is desirable, if not essential \cite{caves, weber}. Quantum dynamics of a test particle and a harmonic oscillator under the influence of linearized GW(s) has been investigated earlier \cite{speli} and recently we have generalized it in a noncommutative (NC) \cite{sny} setting in a couple of papers \cite{sahasg, sahagan} where we treat the theory to first order in the NC parameter.  These studies have revealed the possibility of identifying the NC nature of spacetime (or at least put stringent upper-bounds on the NC parameters \cite{cst, stern, carol, bert0, RB, ani}) through GW detection data. 

Investigations related to the coupling of a background electromagnetic wave with a GW and more specifically, their conjugate effect on test matter, however, has been missing in the literature and is of foremost interest in its own right.
Furthermore, in recent years there is a resurgence of interest in this age-old problem of the coupling of electromagnetic field to GW since experimental investigations are under way with ever increasing precision both in the laboratory or by astronomical observation 
\cite{Tsagas, Marklund}. Also, analytic treatement and numerical simulation of the interaction of a GW with a strongly magnetized plasma have shown that for strong magnetic fields ($\sim 10^{15} {\rm Gauss}$), the GW excites electromagnetic plasma waves where the energy absorbed from the GW by the electromagnetic oscillations is comparable to the energies emitted in the most energetic astrophysical events, such as giant flares on magnetars and possibly even short gamma ray bursts \cite{Isliker}. 


In this paper, we therefore construct the quantum mechanics of a charged test particle gyrating in the presence of a constant background magnetic field along the z-axis, while linearly polarized GW(s) parallel to the magnetic field is passing. 
This system has been analyzed classically by several authors \cite{emgw} 
where one solves the geodesic equation in a linearized GW background.
The problem is worth studying because of its inherent importance in various astrophysical contexts mentioned above.
A quantum mechanical formulation is essential since this would allow us to compare our results with the classical results
and give us deeper insight. Also, since GW couples to electromagnetic waves in plasma mainly by generating electric currents inside the plasma by perturbing the charged particle trajectories \cite{Papa}, the quantum mechanics of the charged particle will also be relevant in that context. Last but not the least, the quantum mechanical formulation would enable us to apply the established framework
of NC electrodynamics \cite{szabo, sw, sch, bcsgas} and gravity \cite{grav, sgfgs, sgrab} to elevate the present study
at the NC level.


In our earlier work \cite{sahasg, sahagan}, we have presented the argument \cite{speli} of working back from the geodesic deviation equation in the proper detector frame to the corresponding Hamiltonian. Following the same path, the Hamiltonian in the present case becomes 
\begin{equation}
{H} = \frac{1}{2m}\left({p}_{j} - \frac{e}{c}A_{j} + m \Gamma^j_{0k} {x}_{k}\right)^2  
\label{e10}
\end{equation}
where the notations have their usual meaning. 
If the GW is propagating along the $z$-axis, due to its transverse nature, ${\Gamma^j}_{0k}$ will have non-zero components only in the ${x}-{y}$ plane. We therefore, confine our attention to that plane. Assuming the symmetric gauge $A_{j} = -\frac{B}{2} \epsilon_{jl}x_{l}$ and using the traceless property of the GW \cite{speli}, the Hamiltonian (\ref{e10}) takes the form
\begin{equation}
{H} = \frac{1}{2m}{p}^{2}_{j} + \frac{1}{2} m \varpi^{2}x^{2}_{l} -\varpi L  + \frac{1}{2}\dot{h}_{jk} x_{k}p
_{j} - \frac{eB}{4 c} \epsilon_{lj} \dot{h}_{jk} x_{l} {x}_{k} ~. 
\label{e11}
\end{equation}
The first two terms represent an ordinary harmonic oscillator with the cyclotron frequency $\varpi = \frac{e B}{2m c}$, the third is the Zeeman term with $L = \epsilon_{lj} x_{l} p_{j}$, the fourth term which is linear in the affine connections, reveals the effect of the passing GW and finally the last term shows the coupling between the GW and the background magnetic field\footnote{Since we are dealing with linearized gravity, a term quadratic in $\Gamma$ has been neglected in eq.(\ref{e11}).}. Note that, in the symmetric gauge we have obtained a Hamiltonian similar to that of a harmonic oscillator along with other terms which can be treated at different orders of perturbation. 
According to the classical treatment when the GW propagates parallel to the magnetic field with a frequency $\varpi^{\prime}$, the coupling of a gyrating particle with the GW becomes very strong  
at twice the cyclotron frequency $\varpi^{\prime} = 2\varpi$ \cite{emgw}. 
We shall try to find whether the quantum mechanical model also show such resonant behaviour. 


Defining raising and lowering operators in terms of the oscillator frequency $\varpi$
\begin{eqnarray}
x_j &=& \left({\hbar\over 2m\varpi}\right)^{1/2}
\left(a_j+a_j^\dagger\right)\>\label{e15a} \\
p_j &=& -i\left({\hbar m\varpi\over 2}\right)^{1/2} 
\left(a_j-a_j^\dagger\right)\>
\label{e15}
\end{eqnarray}
we write the Hamiltonian (\ref{e11}) as 
\begin{eqnarray}
{\hat H} &=& \hbar\varpi\left( a_j^\dagger a_j + 1 \right) + i \hbar\varpi \epsilon_{ij} a^{\dagger}_{i}a_{j} - \frac{i\hbar}{4} \dot h_{jk} 
\left(a_j a_k - a_j^\dagger a_k^\dagger\right)  \nonumber \\
&& - \frac{\hbar}{4} \left[\epsilon_{lj} {\dot h}_{jk}  \left(a_{l}a_{k}  - a_{l}^{\dagger} a_{k}^\dagger \right) \right. \nonumber \\  &&+ \left.  \left(\epsilon_{lj} {\dot h}_{jk} + \epsilon_{kj} {\dot h}_{jl}\right) a_{l}^{\dagger} a_{k}\right].
\label{e16}
\end{eqnarray}
Working in the Heisenberg representation, the time evolution of $a_{i}(t)$ is given by 
\begin{eqnarray}
\frac{da_{i}(t)}{dt} &=& -i{\varpi}a_i + \varpi \epsilon_{ij}a_{j} + \frac{1}{2}\dot h_{ij}a^\dagger_j 
+ \frac{i}{4} \epsilon_{ij} \dot h_{jk} \left(a^{\dagger}_{k} + a_{k}\right) \nonumber\\
&&+ \frac{i}{4} \epsilon_{kj} \dot h_{ji} \left(a^{\dagger}_{k} + a_{k}\right)
\label{e17}
\end{eqnarray}
and that of $a_{i}^{\dagger}(t)$ is the complex conjugate (c.c) of the above equation. 
Next, noting that the raising and lowering operators 
must satisfy the commutation relations
\begin{eqnarray}
\left[a_j(t), a^\dagger_k(t)\right] &=& \delta_{jk}\nonumber \\
\left[a_j(t), a_k(t)\right] &=& 0 = 
\left[a^\dagger_j(t), a^\dagger_k(t)\right]
\label{e18}
\end{eqnarray}
we write them in terms of $a_j(0)$ and $a_j^{\dagger}(0)$, the free operators at time $t=0$, by the time-dependent Bogoliubov transformations
\begin{eqnarray}
a_j(t) &=& u_{jk}(t) a_k(0) + v_{jk}(t)a^\dagger_k(0)\>
\nonumber \\
a_j^\dagger(t) &=& a_k^\dagger(0)\bar u_{kj}(t)  + a_k(0)\bar v_{kj}(t)\>
\label{e19}
\end{eqnarray}
where the bar denotes the c.c and $u_{jk}$ and $v_{jk}$ are the generalised Bogoliubov coefficients. They are $2\times 2$ complex matrices which, due to eq.~$(\ref{e18})$, must satisfy
$uv^{T}=u^{T}v\>,\> u u^\dagger - v v^\dagger = I,$
written in matrix form where $T$ denotes transpose, $\dagger$ denotes c.c transpose and $I$ is the identity
matrix. Since $a_j(t = 0) = a_j(0)$, $u_{jk}(t)$ and $v_{jk}(t)$ have the boundary conditions 
\begin{eqnarray}
u_{jk}(0)& = & I  \quad,\quad  v_{jk}(0) = 0~.
\label{bc0}
\end{eqnarray}
Then, from eq.$(\ref{e17})$ and its c.c, we get the following equations of motions in terms of 
$\zeta = u - v^\dagger$ and $\xi = u + v^\dagger$:
\begin{eqnarray}
\frac{d \xi_{il}}{dt} &=& -i\varpi \zeta_{il} + \varpi \epsilon_{ij} \xi_{jl} + \frac{1}{2}{\dot h}_{ij}\xi_{jl} \>
\label{e21a}\\
\frac{d \zeta_{il}}{dt} &=& -i\varpi \xi_{il} +  \varpi \epsilon_{ij} \zeta_{jl} - \frac{1}{2}{\dot h}_{ij}\zeta_{jl} \nonumber \\
&& + 
\frac{i}{2} \epsilon_{ij} {\dot h}_{jk}\xi_{kl} - \frac{i}{2} {\dot h}_{ij} \epsilon_{jk} \xi_{kl} ~. 
\label{e21b} 
\end{eqnarray}
We shall solve the above eq(s)(\ref{e21a}, \ref{e21b}) for 
the special case of linearly polarized GW(s). 

\noindent In the two-dimensional plane, the GW, 
which is a $2\times 2$ matrix $h_{jk}$, 
is most conveniently written in terms of the Pauli spin matrices as 
\begin{equation}
h_{jk} \left(t\right) = 2f(t) \left(\varepsilon_{\times}\sigma^1_{jk} 
+ \varepsilon_{+}\sigma^3_{jk}\right);\quad \varepsilon_{\times}=\varepsilon_{1},~\varepsilon_{+}=\varepsilon_{3}
\label{e13}
\end{equation}
where $2f(t)$ is the amplitude of 
the GW and $\varepsilon_{\times} \left(t \right)$ 
and $\varepsilon_{+} \left( t \right)$ 
represent the two possible polarization 
states of the GW and satisfy the condition
$\varepsilon_{\times}^2+\varepsilon_{+}^2 = 1$
for all $t$. In case of linearly polarized 
GW(s) however, the polarization states 
are independent of time and $f(t)$ is arbitrary. 
To set a suitable boundary condition we shall assume that 
the GW hits the particle at $t=0$ so that 
\begin{equation}
f(t)=0 \>, \quad {\rm for} \ t \le 0.
\label{bc}
\end{equation}
We now move on to solve eq(s)(\ref{e21a}, \ref{e21b}) 
by noting that any $2\times 2$ complex matrix 
can be written as a linear combination 
of the Pauli spin matrices and identity matrix. 
Hence we make the ansatz : 
\begin{eqnarray}
\zeta_{jk}\left(t \right) &=& A I_{jk} + B_{1}\sigma^{1}_{jk}  
+ B_{2}\sigma^{2}_{jk} + B_{3}\sigma^{3}_{jk} 
\label{form2}\\
\xi_{jk} \left(t \right) &=&  C I_{jk} + D_{1} \sigma^{1}_{jk} 
+ D_{2} \sigma^{2}_{jk} + D_{3} \sigma^{3}_{jk}~.
\label{form1}  
\end{eqnarray} 
Substituting for $h_{jk}$, $\zeta_{jk}$ and $\xi_{jk}$ 
from eq(s)(\ref{e13}, \ref{form2}, \ref{form1}) 
in eq(s)(\ref{e21a}, \ref{e21b}) and comparing the coefficients of 
$I$ and $\sigma$-matrices, we get a set of 
first order differential equations for 
$A, B_{1}, B_{2}, B_{3}, C, D_{1}, D_{2}, D_{3}$:
\begin{eqnarray}
\dot{A} &=& - i \varpi C  +i \varpi B_{2}- \dot{f}\left(\varepsilon_{1}B_{1} 
+ \varepsilon_{3}B_{3}\right) - 2i \dot{f} \left(\varepsilon_{3}D_{1} 
- \varepsilon_{1}D_{3}\right) \nonumber\\
\dot{B}_{1} &=& - i \varpi D_{1} - \varpi B_{3} - \dot{f}\left(\varepsilon_{1} A -
i\varepsilon_{3}B_{2}\right) + 2 \dot{f} \left(\varepsilon_{1}D_{2} 
- i\varepsilon_{3} C \right) \nonumber\\
\dot{B}_{2}  &=& - i \varpi D_{2} + i \varpi A - i \dot{f}\left(\varepsilon_{3}B_{1} - 
\varepsilon_{1}B_{3}\right)-  2 \dot{f} \left(\varepsilon_{1}D_{1} 
+ \varepsilon_{3}D_{3}\right) \nonumber\\
\dot{B}_{3}&=& - i \varpi D_{3} +\varpi B_{1} - \dot{f}\left(\varepsilon_{3} A 
+ i \varepsilon_{1}B_{2}\right) + 2 \dot{f} \left(\varepsilon_{3}D_{2} 
+ i \varepsilon_{1} C\right)\nonumber\\
\dot{C} &=& - i \varpi A  + i \varpi D_{2}+ \dot{f}\left(\varepsilon_{1}D_{1} 
+ \varepsilon_{3} D_{3}\right) \nonumber \\ 
\dot{D}_{1} &=& - i \varpi B_{1} -\varpi D_{3} + \dot{f}\left(\varepsilon_{1}C - i \varepsilon_{3}D_{2}\right) \nonumber \\ 
\dot{D}_{2} &=& - i \varpi B_{2} + i \varpi C + i \dot{f} \left(\varepsilon_{3}D_{1} 
- \varepsilon_{1}D_{3}\right) \nonumber \\ 
\dot{D}_{3} &=& - i \varpi  B_{3} + \varpi D_{1} + \dot{f}\left(\varepsilon_{3} C + i \varepsilon_{1} D_{2}\right) ~.
\label{iteration8} 
\end{eqnarray}
Noting that $|f(t)|<<1$, the above set of equations can be solved
iteratively about its $f(t)=0$ solution by
applying the appropriate boundary conditions (\ref{bc0}, \ref{bc}).
We therefore obtain to first order in the GW amplitude
\begin{eqnarray}
A(t) & = & C(t)= e^{-i\varpi t}\cos(\varpi t)
\label{001}\\
B_{1}(t) & = &  - f(t)e^{-i \varpi t} \left\{\left(\varepsilon_{1} + 2i \varepsilon_{3} \right) \cos \varpi t + \left(\varepsilon_{3} - 2i \varepsilon_{1} \right) \sin \varpi t\right \} \nonumber\\
&& + 3\varpi \left(\varepsilon_{3} - i \varepsilon_{1} \right) \int_{0}^{t} dt^{\prime} f(t^{\prime})e^{-2i \varpi t^{\prime}} \label{002}\\
B_{3}(t) & = &  - f(t)e^{-i \varpi t} \left\{\left(\varepsilon_{3} - 2i \varepsilon_{1} \right) \cos \varpi t - \left(\varepsilon_{1} + 2i \varepsilon_{3} \right) \sin \varpi t\right \} \nonumber\\
&& + 3\varpi \left(\varepsilon_{1} + i \varepsilon_{3} \right) \int_{0}^{t} dt^{\prime} f(t^{\prime})e^{-2i \varpi t^{\prime}} \label{003}\\
B_{2}(t) & = & D_{2}(t) = -ie^{-i\varpi t}\sin(\varpi t)
\label{004}\\
D_{1}(t) & = &   f(t)e^{-i \varpi t} \left(\varepsilon_{1} \cos \varpi t + \varepsilon_{3} \sin \varpi t\right) - \varpi \left(\varepsilon_{3} - i \varepsilon_{1} \right) \int_{0}^{t} dt^{\prime} f(t^{\prime})e^{-2i \varpi t^{\prime}} \label{005}\\
D_{3}(t) & = &   f(t)e^{-i \varpi t} \left(\varepsilon_{3} \cos \varpi t - \varepsilon_{1} \sin \varpi t\right) + \varpi \left(\varepsilon_{1} + i \varepsilon_{3} \right) \int_{0}^{t} dt^{\prime} f(t^{\prime})e^{-2i \varpi t^{\prime}}. \label{006}
\end{eqnarray}

\noindent Computing the above expressions for $A, B_{1}, B_{2}, B_{3}, C, D_{1}, D_{2}, D_{3}$ 
assuming a monochromatic GW waveform $f(t)=f_{0}e^{i\varpi^{\prime}t}$ oscillating with frequency $\varpi^{\prime}$ 
finally lead (by carrying out the procedure in \cite{sahasg, sahagan}) to the following general expression of 
$\langle X_{1}(t)\rangle$ 
\begin{eqnarray}
\langle X_{1}\left(t\right)\rangle & = & \sqrt{\frac{2\hbar}{m \varpi}} {\mathrm{Re}}[a_{1}\left(t\right)] = X_{1}(0) \cos^{2} \varpi t  + P_{1}(0) \frac{\sin 2 \varpi t}{2 m\varpi} + X_{2}(0) \frac{\sin 2\varpi t}{2}  + P_{2}(0) \frac{\sin^{2} \varpi t}{m\varpi} \nonumber\\
&& + X_{1}(0)f_0 \left[\frac{\varepsilon_{1}}{4}F +  \varepsilon_{3}\left\{ \frac{1}{2} \cos{\varpi^{\prime} t} + G \right\}\right] 
+ X_{2}(0)f_{0}\left[\varepsilon_{1}\left\{ \frac{1}{2} \cos{\varpi^{\prime} t} + G\right\} - \frac{\varepsilon_{3}}{4}F\right] \nonumber\\
&& + \frac{P_{1}(0)}{m \varpi}f_{0}\left[\varepsilon_{1} \left\{-\frac{1}{2} \cos{\varpi^{\prime} t} + G\right\} 
- \frac{\varepsilon_{3}}{4} F \right] -\frac{P_{2}(0)}{m \varpi}f_{0}\left[ \frac{\varepsilon_{1}}{4} F + \varepsilon_{3} \left\{-\frac{1}{2} \cos {\varpi^{\prime} t} + G\right\} \right]\nonumber\\
\label{x1}
\end{eqnarray}
where $\langle\vec{R}_{0}\rangle = \left(X_{1}(0), X_{2}(0)\right)$ and $\langle\vec{P}_{0}\rangle = \left(P_{1}(0), P_{2}(0)\right)$ 
are the initial position and momentum expectation values when the GW just hits the system at time $t=0$ and
\begin{eqnarray}
F \left( t \right) & = &  - \frac{\varpi^{\prime}}{2 \varpi + \varpi^{\prime}} \sin {\left(2 \varpi + \varpi^{\prime}\right)t} + \frac{\varpi^{\prime}}{2 \varpi - \varpi^{\prime}} \sin {\left(2 \varpi - \varpi^{\prime}\right)t} \nonumber\\
G \left( t \right) & = & \frac{2 \varpi^{2}}{4\varpi^{2} - \varpi^{\prime 2}} + \frac{\varpi^{\prime}}{4(2 \varpi + \varpi^{\prime})} \cos {\left(2 \varpi + \varpi^{\prime}\right)t} - \frac{\varpi^{\prime}}{4(2\varpi - \varpi^{\prime})} \cos {\left(2 \varpi - \varpi^{\prime}\right)t}~.
\end{eqnarray}
The above result is very similar to the one obtained earlier in \cite{sahagan} 
in noncommutative two-dimensional space and
clearly shows that some of the oscillatory terms representing the response of the
particle in the background electromagnetic field to the GW, will exhibit resonance when the condition
$\varpi^{\prime}=2\varpi$ is satisfied. 
This is indeed reassuring that the condition of resonance obtained through 
our quantum mechanical analysis is in conformity with the classical analysis \cite{emgw}.
Hence, one can in principle detect the GW and find its frequency $\varpi^{\prime}$
by tuning the harmonic oscillator frequency $\varpi$ (which in turn depends on the applied magnetic field $B$) till
resonance is achieved. Realistic scenarios can however be obtained if various forms of periodic GW signals
with more than one frequency are used to carry out similar computations. We hope to report this in future.


\end{document}